# Motility-dependent selective transport of active matter in trap arrays: Separation methods based on trapping-detrapping and deterministic lateral displacement

Vyacheslav R. Misko[1,2,*], Franco Nori[2,3,4] and Wim De Malsche[1]

[1] µFlow group, Department of Chemical Engineering, Vrije Universiteit Brussel, Pleinlaan 2, 1050 Brussels, Belgium
[2] Theoretical Quantum Physics Laboratory, Cluster for Pioneering Research, RIKEN, Wakoshi, Saitama 351-0198, Japan
[3] Quantum Computing Center, RIKEN, Wakoshi, Saitama, 351-0198, Japan
[4] Physics Department, University of Michigan, Ann Arbor, Michigan 48109-1040, USA

Selecting active matter based on its motility represents a challenging task, as it requires different approaches than common separation techniques intended for separation based on, e.g., size, shape, density, and flexibility. This motility-based selection is important for, e.g., selecting biological species, such as bacteria or highly motile sperm cells for medically assisted reproduction. Common separation techniques are not applicable for separating species based on motility as such species can have indistinguishable physical properties, i.e., size, shape, density, and differ only by their ability to execute self-propelled motion as, e.g., motile, and immotile sperm cells. Therefore, selecting active species based on motility requires completely different approaches. Some of these have been developed including sperm cell selection techniques, e.g., swim-up techniques, passive selection methods based on the ability of highly motile sperm cells to swim across streamlines, as well as more sophisticated techniques. Here we theoretically demonstrate via numerical simulations various efficient methods of selection and separation based on the motility of active species using arrays of traps. Two approaches are proposed: one allowed the selective escape of motile species from traps, and the other one relying on a deterministic lateral displacement (DLD)-type method. As a model system, we consider self-propelled Janus particles whose motility can be tuned. The resulted separation methods are applicable for separation of biological motile species, such as bacteria or sperm cells, as well as for Janus micro- and nanoparticles.

## 1. Introduction

The term "active matter" refers to various system of particles, living cells or other species that can execute self-propelled motion, i.e., move on their own "engine", without any external driving. This self-propelled motion results from the conversion of the consumed energy (e.g., energy due to chemical reactions) to the mechanical energy providing the motion. Such systems are out of thermodynamic equilibrium as they require a constant energy supply to support their activity. Examples of self-propelled particles are chemically fuelled Janus particles [1-10] that are usually half-covered on one of the hemispheres by a thin layer of gold or platinum that catalyses a chemical reaction (hydrogen peroxide decomposition) leading to the surface flow of ions that in turn induces self-propulsion. New-generation high-motility Janus particles can also propel in pure water [11, 12] making them bio-compatible and thus suitable for bio-applications, e.g., for bio-sensing.

It can be desirable to select the highest motility Janus particles out of an ensemble of: (i) many Janus particles, whose self-velocity can vary due to, e.g., fluctuations of the size of the metal cover layer, etc.; and (ii) immotile particles present in the mixture. High motility Janus nanoparticles are needed for various applications, e.g., for mixing in microfluidics or for the elimination of drug-resistant biofilms [13, 14]. In general, for any experimental purposes it is always desirable to select high-motility Janus particles, including for the purpose of using these as a model system for motile cells or bio-particles.

In case of living motile cells, their ability to move is of vital importance, and nature can select these according to their motility. Indeed, sperm motility plays a crucial role in reproduction. Only the sperm cell that passed the stringent motility selection (and other types of selection) in the reproductive tract can reach and fertilize the oocyte. Therefore, medically assisted reproduction (MAR) implies, in the first place, selection of motile sperm cells that can be further used for either intra-uterine insemination, in vitro fertilization, or for intracytoplasmic sperm injection [15]. Examples of commonly used sperm preparation techniques are swim-up and density gradient centrifugation [16]. More advanced selection methods are based on sophisticated morphological assessment, electrical charge, and molecular binding. These methods mimic in part selection mechanisms in a female reproductive tract [16]. Various microfluidic setups have been proposed to select high-motility sperm cells including a passively driven self-contained integrated microfluidic device for separation of motile sperm [17]. The principle of this device is based on the ability of motile sperm cells to cross streamlines in a laminar fluid stream [17-19]. Further modification and integration of passive devices allowed several sperm-selecting devices, like a "sperm syringe" [20] and other separation devices [21, 22].

Other microfluidic methods of motile sperm selection do not require external driving and use the activity of motile sperm cells themselves for selecting these from immotile cells and debris. These methods are based on the ability of motile sperm cells to accumulate in the corners of asymmetric obstacles [23] or propel in a preferrable direction through microchannels [24-26] or in asymmetric obstacles [27], where motile sperm cells reveal the effect of self-rectification of active motion [6] (which is of different nature than rectification in driven dynamics [28, 29]). The motility of sperm cell (or synthetic microswimmers like Janus particles) provides their selective escape from, e.g., a circular domain [30, 31], that can also be used for selecting self-propelled particles or motile cells. In addition, alternative methods have been proposed that allowed to selectively enhance the motility of sperm cells. Thus, sperm carrying micromotors [32] were able to capture, transport, and release a sperm cell in fluidic channels, and deliver it to the oocyte for fertilization. Another method is based on the effect called "motility transfer" [33] where a more active guest species in a binary mixture of active swimmers transfers its motility to a less active host species (e.g., sperm cell). As guest species, high-motility artificial microswimmers, i.e., bio-compatible catalytic Janus particles [34] can be used.

The ability of motile cells or self-propelled particles to escape from potential-energy traps is a distinguishing feature that allows to select them from immotile cells or particles that only perform Brownian motion. This method has been recently demonstrated in experiments [35] selecting sperm cells based on their motility in an acousto-fluidic device in a continuous flow regime. There, only highly motile sperm cells were capable of escaping from an acoustic trap and be collected in the side co-flows, while immotile and weakly motile cells and debris remained trapped and transported in the central flow [35].

In this work, we explore the ability of motile particles to selectively escape from potential-energy traps, for their separation from immotile species. Using numerical simulations, we demonstrate two methods, one discontinuous, where motile particles can selectively escape from traps in the presence of external weak driving, while passive species remain trapped. The other proposed method uses a deterministic lateral displacement (DLD)-type array of traps where both, motile and immotile, species escape the traps under the action of the external driving. In this approach, the passive species preferably follow the rows of the array (i.e., the displacement DLD mode) while the motile particles preferably move along the direction of the driving flow (i.e., the zigzag DLD mode). This difference in behaviour provides an effective separation of motile particles from the immotile species. Note that the proposed DLD-type device essentially differs from traditional DLD devices where pillars, which are repulsive obstacles, are used to guide the particles, while in our case the "obstacles" are represented by *attractive* potential-energy traps, which leads to a rather non-trivial trapping-detrapping dynamical behaviour [36].

**2. An active particle in a trap**

Let us first consider a self-propelled particle characterized by a self-propulsion velocity $v_0$ moving inside a parabolic trap of strength $A$. The force exerted on the particle due to the trap is then:

$$\mathbf{F}_p = -A\mathbf{r}, \tag{1}$$

where $\mathbf{r}$ is the vector position of the particle. According to the theoretical model developed in [37], in an infinite harmonic trap, for a single active particle, the stochastic equations of motion for the $i$-th self-propelled particle can be written as:

$$\frac{dr_i}{dt} = -Ar_i + \cos\theta_i, \tag{2}$$

$$\frac{d\theta_i}{dt} = \sqrt{2D_r}\eta(t) - \frac{v_0}{r_i}\sin\theta_i, \tag{3}$$

where the angle $q_i$ is between the direction of motion of the particle and the normal to the trap, $h(t)$ is a Gaussian unit white noise, and $D_r$ is the rotational diffusion coefficient. Using dimensionless variables $t = tD_r$, $r_i = r_iA/v_0$, and $h(t) = h(t)/\sqrt{D_r}$, the equations (2) and (3) can be rewritten as:

$$\frac{dr_i}{dt} = -\frac{A}{D_r}(\rho_i - \cos\theta_i), \tag{4}$$

$$\frac{d\theta_i}{dt} = \sqrt{2}\eta(\tau) - \frac{A}{\rho_i D_r}\sin\theta_i. \tag{5}$$

Thus, the angle $q_i$ undergoes rotational diffusion in an effective potential $(A/r_i D_r)(1 - \cos q_i)$, whose amplitude diverges as $D_r/A \to 0$. In this limit, as shown in Ref. [37], the particle is almost always at the border of the trap, although the fluctuations of the angle $q_i$ prevent it from reaching exactly $r_i \sim v_0/A$, in a band of thickness $D_r v_0/2A^2$.

The above model is valid for an infinite harmonic trap [37]. We adopt this model for a truncated parabolic trap that is cut off at a radius $R = D/2$. Therefore, an active particle will be trapped only when $v_0/A < R$, and will be untrapped otherwise.

The dynamics of active and passive particles in finite parabolic traps has been earlier analysed in [36], where it was shown that active particles can assist trapping and detrapping of passive particles in a binary mixture of active and passive particles. Here we are interested in the selective detrapping of active particles from the traps in the presence of an additional driving force, i.e., a flow. In the presence of a flow, particles are driven by the Stokes drag force,

$$\mathbf{F}_d = -6p\, h_f\, a\, \mathbf{v}, \qquad (6)$$

where $h_f$ is the viscosity of the fluid, $a$ is the radius of the particle, and $\mathbf{v}$ is the particle velocity.

As follows from equations (4) and (5) and the analysis of Ref. [36], the non-zero self-velocity $v_0$ of active particles makes the "residence" of self-propelled particles strikingly different from that of passive particles. While passive particles relax, in the absence of external driving, toward the bottom of the parabolic trap (provided the thermal fluctuations are weak), self-propelled particles prefer to "reside" at the walls of the parabolic trap. The self-propelling velocity $v_0$ prevents them from "falling" to the bottom of the trap. Consequently, these active particles can be extracted *easier* from the parabolic trap than passive particles. Indeed, even from a simple energy consideration it is clear that in order to extract a passive particle from the bottom of the trap, an energy equivalent to its potential energy, $E = AR^2/2$, is required.

On the other hand, an active self-propelled particle possesses a potential energy $E_{sp} = AR_p^2/2$ in the trap, where $R_p \sim v_0/A$ is the radius of its precession inside the parabolic trap, and the energy to extract such active particle is $DE = A(R^2 - R_p^2)/2$. Therefore, if the self-velocity is large enough to keep the self-propelled particle near the escape point of the trap (but insufficient to escape, i.e., $R_p < R$), adding a weak additional force would facilitate the escape event. This phenomenon resembles stochastic resonance, where a small random perturbation can trigger a transition in a system that in a state close to the transition point [38].

The principle of external flow-assisted escape of active particles from a trap is further illustrated in Fig. 1.

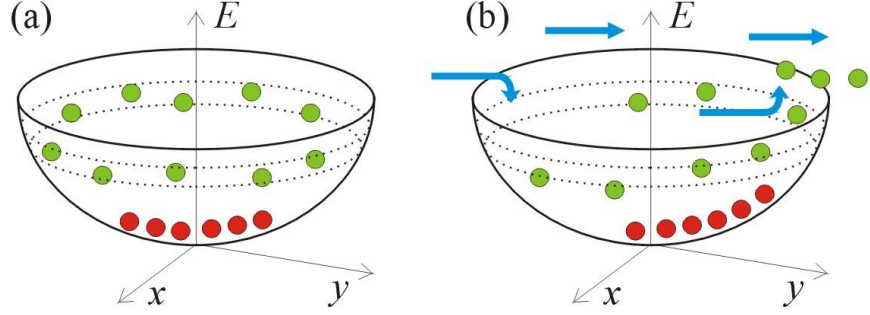

**Fig. 1.** Flow-assisted escape of active particles from a parabolic trap. (a) Distribution of passive (shown by red circles) and active (green circles) particles inside a parabolic trap. Passive particles reside near the bottom of the trap while self-propelled active particles reside near the trap edges away from the bottom, inside a band of thickness $D_r v_0/2A^2$. (b) The drag force exerted on the particles due to a flow (flow direction indicated by blue arrows) can cause their escape. A weak force facilitates the escape of active particles, while passive particles remain trapped closer to the bottom (see Eqs. (4), (5) and Refs. [36, 37]).

As it is clear from Fig. 1, the different species, i.e., self-propelled particles and passive beads, become already separated inside the trap. Indeed, when a mixture of them enters the trap, the initially mixed particles of different sorts become separated: the passive species is collected near the bottom (i.e., near the center of the trap) while the active particles are dynamically located near the edge of the trap. Therefore, these active particles can be extracted easier from a trap, compared to the passive particles. This difference in the behaviour of active and passive particles inside the trap is used for the first of the separation methods discussed here, i.e., for the selective detrapping of active particles (or motile cells) from arrays of traps. The connection between the particle preselection inside the trap and the second separation method, i.e., in the DLD-type device, is not that straightforward, although the difference in the location of the active and passive particles inside the trap is also crucial for that separation mechanism, as will be discussed below.

## 3. Trapping-detrapping mechanism of motile particle separation in arrays of traps

The principle of trapping-detrapping particle selection based on their motility in an array of traps is schematically presented in Fig. 2.

In the previous section, we analysed the motion of self-propelled particles, characterized by the self-velocity $v_0$, in parabolic traps and estimated a single-particle solution for the radius of precession of a self-propelled particle as a function of the strength of the trap, $A$, as well as the band thickness where the particles is located inside the trap.

Note that these estimates were found for a single point-like particle. Now we consider a more complex system of total $N$ particles of a finite radius, $a$, including self-propelled particles moving with self-velocity $v_0$, and passive beads; all of these in the presence of thermal noise. In this way, thermal diffusion is taken in account. In addition, an external flow with velocity, $v_{\text{flow}} \equiv v_f$, is applied that drives the system over an array of parabolic traps. In simulations, an array of 4 x 4 traps is embedded in a square simulation region where the boundary conditions at two opposite boundaries are open (inlet and outlet) and two others (channels sides) are reflecting boundaries, in the overdamped regime. To describe the dynamics of this complex system, we numerically solve a set of $N$ Langevin-type equations of motion for all the particles. The equations of motion and the details of the simulations are presented in the "Methods" section.

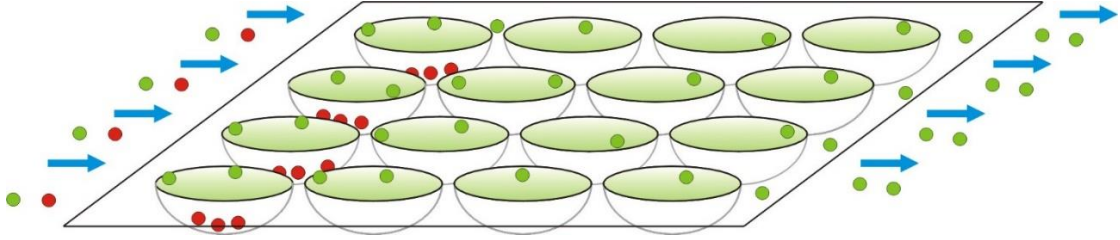

**Fig. 2.** Flow-assisted trapping-detrapping mechanism of separation of active particles driven via an array of traps. A mixture of active particles (motile cells) (shown by green circles) and passive beads (immotile cells) (shown by red circles) is infused into the separation device from the left side. Both the species, active and passive, are trapped by the potential energy traps (and in the absence of the additional flow they would remain trapped). Due to the additional (weak) flow (flow direction indicated by blue arrows), the active species escape the traps and move towards the right side of the device, where they can be separately collected. The passive particles remain trapped although, depending on the applied flow rate, can also be detrapped and trapped by other traps.

The trajectories of passive and active particles infused in a device with an array of parabolic traps are shown in Fig. 3 by red and green lines, correspondingly. All the particles of the mixture are infused from the left side of the device. Passive particles undergo Brownian motion due to thermal fluctuations characterized by (dimensionless) temperature $T = 2.5 \times 10^{-3}$ and rotational diffusion coefficient $D_r = 2.5 \times 10^{-2}$. These thermal diffusion parameters are used throughout this work. The typical parameters of the parabolic potential-energy traps considered here are as follows: trap radius, $R = 3.5$, and tramp strength, $A = 1/2R$. The traps are located at the centers of 4 x 4 square units of 7.5 x 7.5 arranged in a 30 x 30 square matrix.

When the particles of both sorts reach the array of traps, these become trapped in the first row of traps. Located on the left side. We note that, according to the analytical results presented at the beginning of Sec. 2, passive particles (their trajectories shown by red lines) accumulate near the bottom (center) of the traps, while active particles perform nearly circular motion close to the edges of the traps. These active particles therefore can easier overcome the potential-energy barrier and escape the trap.

After the escape, these particles become trapped again in the next row, as shown in Fig. 3, and finally, after repeated events of trapping and detrapping, these arrive at the right-side chamber of the device, where these can be collected separately from the passive particles. In this way, a discontinuous trapping-detrapping particle separation mechanism in an array of potential traps is demonstrated in this device. Passive particles remain trapped in the traps, and the device requires some resetting in order to remove those passive particles from the traps.

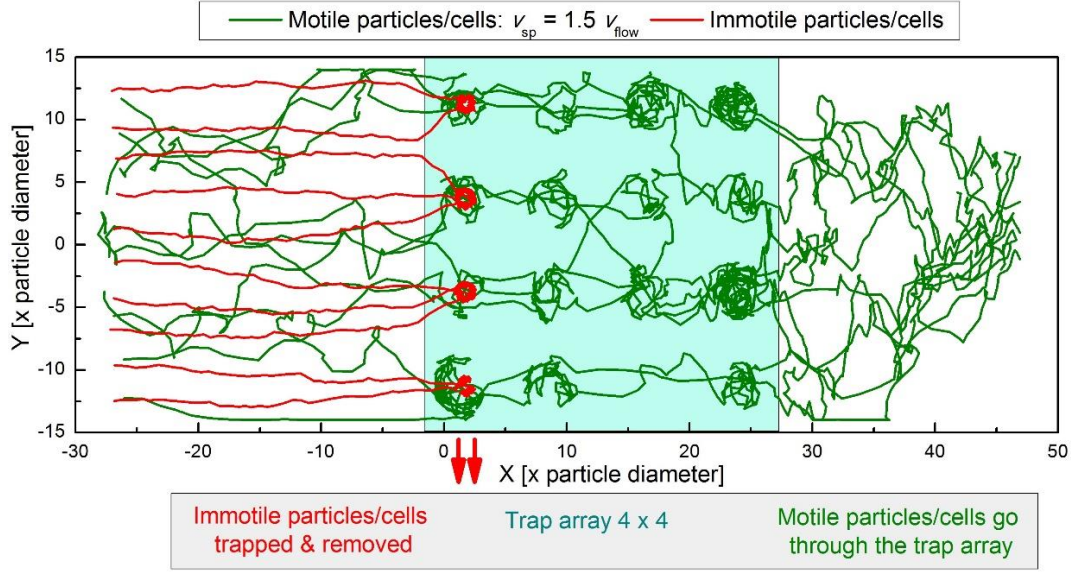

**Fig. 3**. Selective trapping of particles depending on their motility. The trajectories of passive particles (executing Brownian motion) are shown by red colour, and the trajectories of active particles (Janus particles or motile cells) characterized by self-velocity, $v_0 \equiv v_{sp}$, are shown by green colour. A mixture of passive and active particles is infused from the left side. It is driven by a transport flow with velocity $v_{flow}$, which is chosen as the unit of velocity. Correspondingly, the self-propelled velocity is $v_0 = 1.5 v_{flow}$. Both species, active and passive, are trapped in the first (left) row of traps, and active particles escape the traps and, through a series of trapping-detrapping events, reach the right-side chamber of the device and are collected separately. Passive particles remain trapped. The thermal diffusion parameters are: temperature $T = 2.5 \times 10^{-3}$ and rotational diffusion coefficient $D_r = 2.5 \times 10^{-2}$.

Not only active particles (or motile cells) can be separated from immotile particles (cells) using this device. We can imagine, e.g., the need of separating high-motility cells, like sperm cells, from lower motility cells. This task is of importance, e.g., for the purpose of MAR, as mentioned above. To demonstrate separation of high-motility cells (or particles) from low-motility cells (particles), we infuse a mixture of motile cells characterized by self-velocities $v_0 = 1.5 v_{flow}$ ("high motility") and $v_0 = v_{flow}$ ("low motility") and repeat the same numerical experiment as presented above in Fig. 3. The results are shown in Fig. 4. Again, both species follow the transport flow, reach the array of traps, and become trapped in the first (leftmost in the figure) row of the traps. However, due to their nonzero self-velocity, $v_0$, both species are capable of escaping the traps, with different escape probability: higher for the high-motility particles (cells) and lower for low-motility particles (cells). As a result, for a given set of parameters, the low-motility particles become trapped in successive rows of traps, as shown in Fig. 4 (their trajectories are shown by magenta colour). In contrast, the high-motility particles eventually escape all the traps, and after the trapping-detrapping events they reach the right-side chamber, where they are collected separately.

Due to the non-zero probability of the escape of low-motility species (and not only the high-motility species), for long enough times, all of these low-motility particles should also escape from the trap. But this consideration in general is not relevant for self-propelled particles, as self-propelling itself is a "transient" effect. In the long-time limit, $t \to \infty$, diffusive properties of self-propelled particles and Brownian particles become indistinguishable as self-propulsion displays itself only for finite times (which can be very long, e.g., sperm cells can remain motile up to few days after delivering them in the female reproductive tract, and synthetic Janus particles show appreciable activity up to few months). Therefore, the results shown in Fig. 4 should be understood as valid for finite times, where the separation effect occurs for times much shorter than the observation time. In other words, the separation process occurs *faster* (e.g., in seconds) than the time needed for the low-motility particles to eventually escape the traps (e.g., in minutes or hours). Then the device should be reset, to be ready for a new separation cycle.

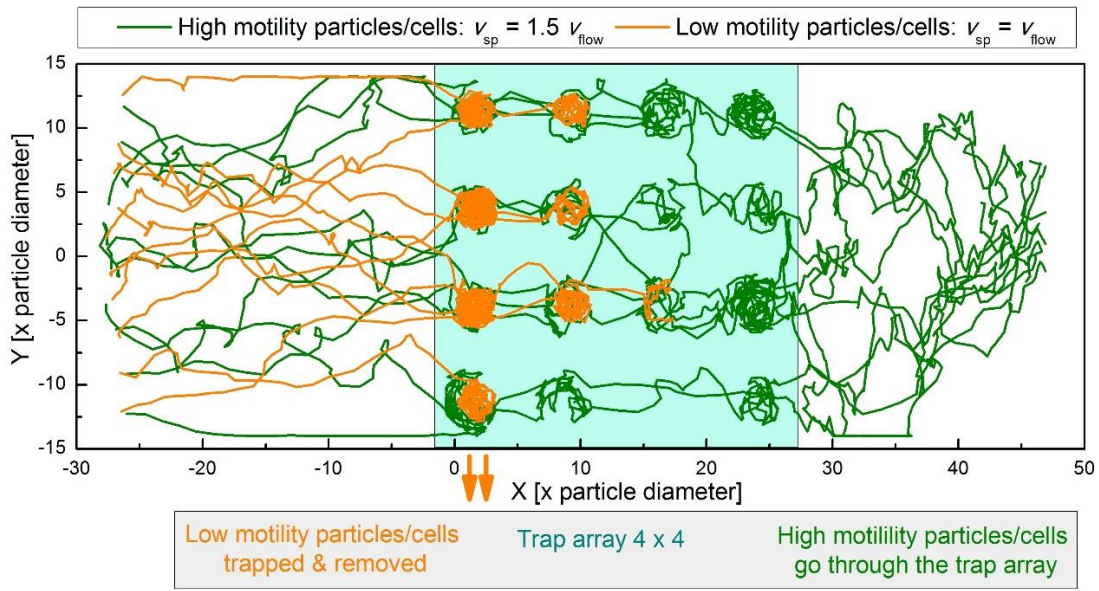

**Fig. 4.** Selective trapping of particles depending on their motility. Same as in Fig. 3, but for a mixture of low-motility and high-motility particles (cells). The trajectories of low-motility particles with self-propelled velocity $v_0 = v_{\text{flow}}$ are shown by orange colour, and the trajectories of high-motility particles with self-propelled velocity $v_0 = 1.5 v_{\text{flow}}$ are shown by green colour. The mixture of particles is infused from the left side of the channel. Both species undergo trapping-detrapping motion via the array, but the high-motility particles (cells) reach the right-side chamber (during the observation time), while low-motility particles remain trapped. The thermal diffusion parameters are the same as in Fig. 3.

We note that the presented separation method of motile particles based on the trapping-detrapping mechanism, as described above, is very robust. It allows to separate motile particles from immotile (or high motility from low motility) in a broad range of parameters including particle motility, trap strength (and size), and flow strength.

# 4. Motile particle separation in arrays of traps using a deterministic lateral displacement mechanism

In the previous subsection, we discussed a discontinuous method of particle (cell) separation based on their motility. This means that, after each cycle of separation of some number of particles or cells, the device requires resetting, after which the separation can be continued. This limitation of the method is not essential when only a very limited amount of sample should be separated. For example, one may need only a very small number of high-motility sperm cells for the need of MAR. However, for broader applications, i.e., technological applications, where larger amounts of particles need to be separated, continuous methods of separation are required. In case of active and passive particles, a large amount of high-motility Janus particles may be required, e.g., for the elimination of drug-resistant biofilms [13, 14].

In this subsection, we discuss a continuous method of motile particle selection and separation, using an array of potential-energy traps. The basic idea of the method is sketched in Fig. 5.

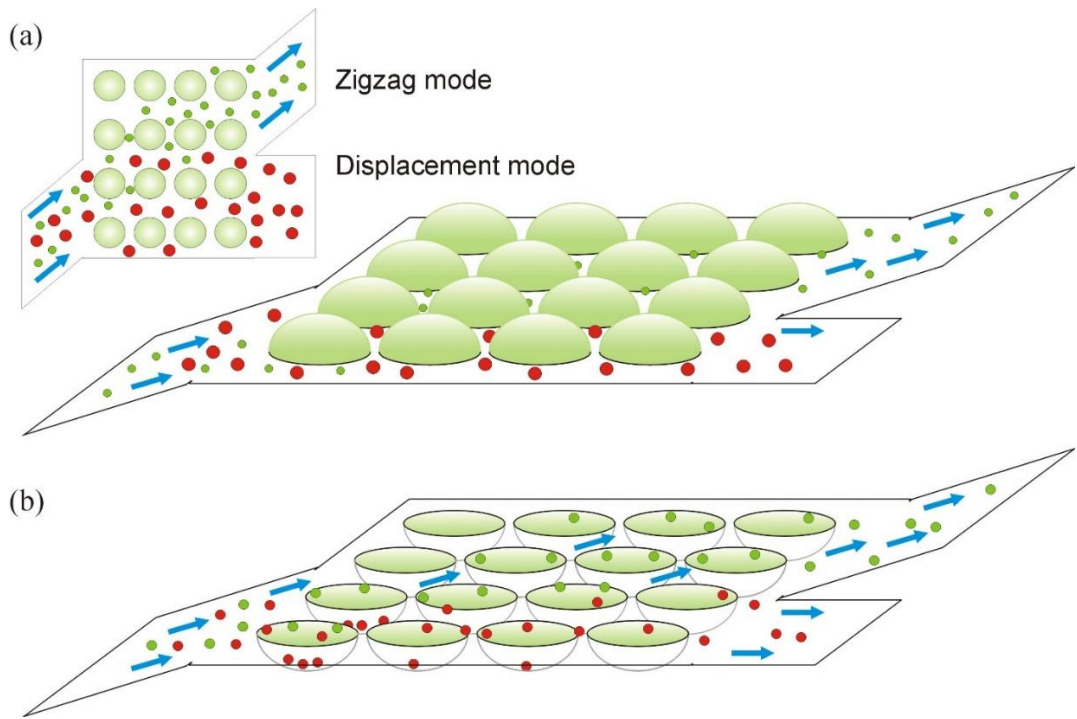

**Fig. 5.** Deterministic lateral displacement (DLD) device for separation of particles depending on their motility. (a) Conventional DLD device for particles separation, i.e., large (large red circles) from small (small green circles). The inset shows the top view of the DLD device. A mixture of particles is infused from the left bottom corner. Large particles are guided by rows of repulsive obstacles (pillars) ("displacement" mode), while small particles, that can go through the horizontal gaps between the pillars, follow the direction of the flow ("zigzag" mode). The large and small particles are collected separately. (b) A sketch of a DLD device with arrays of attractive potentials for particles separation based on their motility. Both the species, motile (green circles) and immotile (red species) can escape the traps. Immotile particles become trapped in the closest adjacent row and move along that row ("displacement" mode), while motile particles have random direction of motion and can move to other rows. Between the traps, they are guided by the flow ("zigzag" mode).

As illustrated in Fig. 5, the principle of particle separation based on their motility in a DLD-type device is rather different from separation of particles, e.g., large versus small, in a conventional DLD device. The principle of a conventional DLD device (Fig. 5a) implies that one type of particles (small particles) can go through the gaps between the different rows of the DLD array (in the horizontal direction in Fig. 5a) as well as through *smaller* gaps between the traps in the same row. Therefore, on average, these can follow the direction of the driving flow, i.e., by taking few steps in one direction (i.e., horizontal) and then a step in the other direction (vertical). In this way, a "zigzag" mode is realized. On the contrary, large particles can go only through the large gaps between the pillar rows in the vertical direction (inset of Fig. 5a), i.e., they move along the horizontal direction *guided* by the pillar rows, thus realizing the "displacement" mode of motion in the DLD device. Somewhat similar dynamics can be found in other systems of particles undergoing successive trapping and detrapping transitions [39] when varying the angle of injection of particles in substrates with arrays of traps.

A recent example of successful application of a (conventional) DLD device is presented in Ref. [40]. The authors presented improved DLD arrays allowing efficient focusing of particles and higher particle concentration enhancement. Thus, their 18 μm-gap device showed 11-fold enrichment of 7 μm particles, and more than 50-fold for 10 μm particles and Jurkat cells [40].

It is clear that this principle, in general, cannot provide separation of particles based on their motility, e.g., of particles or cells of the same size. Therefore, we propose to employ arrays of *attractive* traps. The principle of separation of this DLD device can be understood as follows. We assume that both types of particles, motile and immotile, can be trapped and can escape the traps. When escaping the trap, immotile particles leave the trap at the side of the trap opposite to the point where they entered the trap, following the largest component (which is horizontal for the infusion angles < 45°) and with the *highest probability* these particles become re-trapped by the closest-neighbour traps in the same row. This sequential trapping-detrapping process guides the particles in the direction along the row of traps (Fig. 5b). Active particles (or motile cells), due to their own self-propulsion, perform rotational motion inside the traps (as discussed above). Therefore, at the moment of escape they can move in *any* direction. This escape point can be other than that near the closest trap (as in case of passive particles). After escaping from the trap, these active particles become driven along the direction of the driving flow. In this way, the "zigzag" mode is realized in this DLD device with attractive particles. We note that repulsive and attractive obstacles have been earlier employed for particle separation in model simulations in Ref. [29]. The obstacles were arranged in arrays and provided transverse rectification of motion of particles driven by an AC driving. The effect was due to the triangular shape of the obstacles.

Alternatively, the mechanism of separation in a DLD device with attractive sites can be understood in terms of potential energy minimization. When moving through an array of potential traps, passive particles follow the minimum energy landscape, so during each event of trapping these passive particles reach the minimum of the potential energy at the bottom of the trap. Between these energy minima, the passive particles need to overcome the energy barriers formed by the gaps between the adjacent traps. It is clear that the smallest gaps between the adjacent traps minimize the average potential energy during this motion. On the contrary, active particles always remain close to the edge of the trap, and in this way their potential energy does not change much during trapping events as compared to that when moving outside the traps. In other words, the motion of these active particles is less impacted by the traps, provided their self-velocity is high enough. Therefore, on average they follow the direction of the driving flow, and (for strong enough drive or weak enough traps) the trapping-detrapping events can be considered as a *perturbation*. This mode is the "zigzag" mode in this DLD device when the active particles jump either to the closest traps or to other traps in the direction of the flow. It is worth noting that this difference in the motion of active and passive particles has a statistical nature (thus implying many-particle system), and not deterministic as in the case of the conventional DLD devices when, e.g., large particles cannot go through the small gaps. Here, there is a non-zero probability of active particles moving along the trap row, as well as of sudden jumps of passive particles between the traps in a neighbouring row (e.g., for angles close to 45°). Similar stable diagonal trajectories or stable lock-in dynamical states have been found in [39]. Therefore, for better separation, large trap arrays should be employed, and the DLD infusion angle properly optimized.

Considering these conditions, we have performed simulations of the transport of active particles, with varying self-propelling velocity, through an array of potential-energy traps, for different flow directions with respect to the direction of the trap arrays. Further, we simulated the transport of a mixture of active and passive particles, to analyse the separation based on motility of the particles. Some representative simulation results for active particles are shown in Fig. 6 and Fig. 7.

Figure 6 illustrate the impact of the self-velocity of active particles on the ability of the particles to either (preferably) follow the direction of the trap rows ("displacement" mode, Fig. 6a) or of the driving flow ("zigzag" mode, Fig. 6b), for a given angle between the direction of the applied driving and the trap row direction, i.e., g = 14°. Thus, for the lower self-propelling velocity, $v_0 = 0.7 v_{flow}$ (Fig. 6a), the active particles preferably move along the two lower rows of the traps (inside their trap array entrance gap on the left side of the DLD device). This behaviour is similar to what expected for low-motility or immotile particles, as discussed above. For higher self-propelling velocity, $v_0 = v_{flow}$ (Fig. 6b), the dynamical pattern substantially changes: the trajectories show that active particles preferably jump to the traps in the neighbour rows (along the vertical direction) and escape the DLD device mainly through the two upper rows, demonstrating the "zigzag" mode of propulsion. Thus, we showed that by increasing the self-propelling velocity, we can effectively switch between the (preferably) "displacement" and "zigzag" modes, which is indicative of the possibility to distinguish and separate, e.g., low-motility from high-motility active particles (or motile cells like sperm cells).

Next, the impact of the self-velocity on the dynamics of active particles is presented in Fig. 7, for a larger angle between the direction of the applied driving and the trap row direction. For a larger angle between the flow direction and that of the trap rows, g = 21.8°, and low self-velocity, $v_0 = 0.25 v_{flow}$, most of the particles leave the array through the two middle rows (unlike in the case of $v_0 = 0.7 v_{flow}$ and g = 14° shown in Fig. 6a, where the particles mainly followed the trap rows). This trend remains valid for a higher self-velocity, $v_0 = 0.5v_{flow}$, although the dispersion in the escape location of the particles from the array increases (Fig. 7b). Thus, a higher self-velocity is required to achieve a more pronounced "zigzag"-mode regime (like in the case $v_0 = v_{flow}$ shown in Fig. 6b), even for larger driving flow angles.

(a)

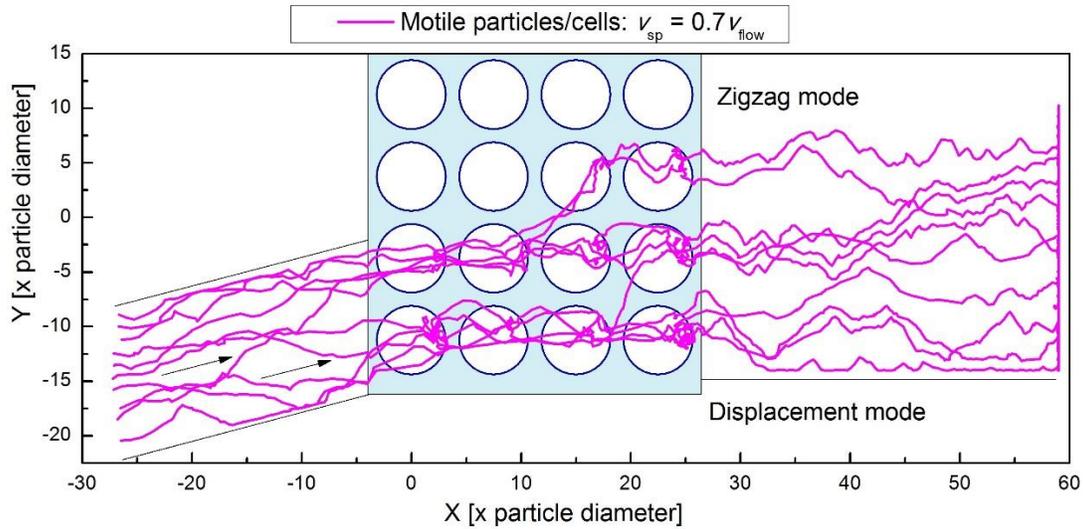

(b)

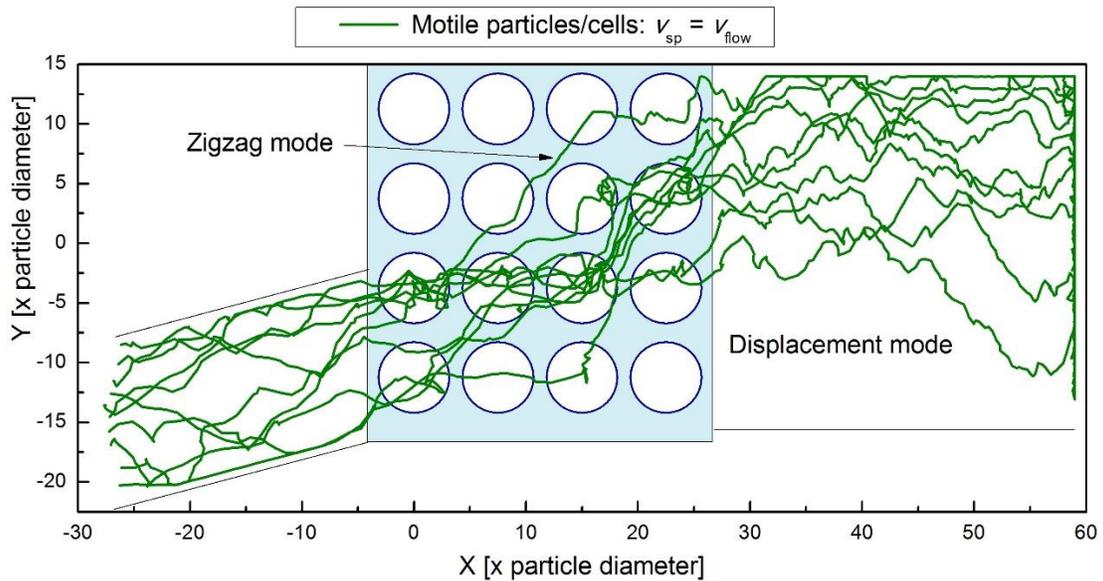

**Fig. 6.** Trajectories of self-propelled particles driven by an external flow with velocity $v_{flow} = 1$ through a 4 x 4 array of attractive parabolic potential energy traps, for a small angle, $\gamma = 14°$, between the direction of the flow and the array row, and for the values of the self-velocity: $v_0 = 0.7v_{flow}$ (a) and $v_0 = v_{flow}$ (b).

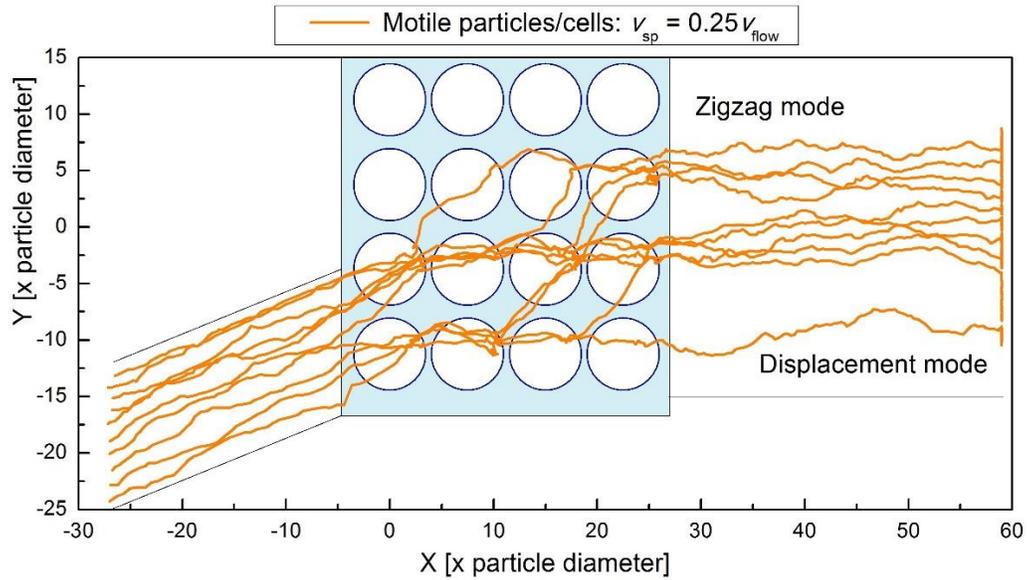

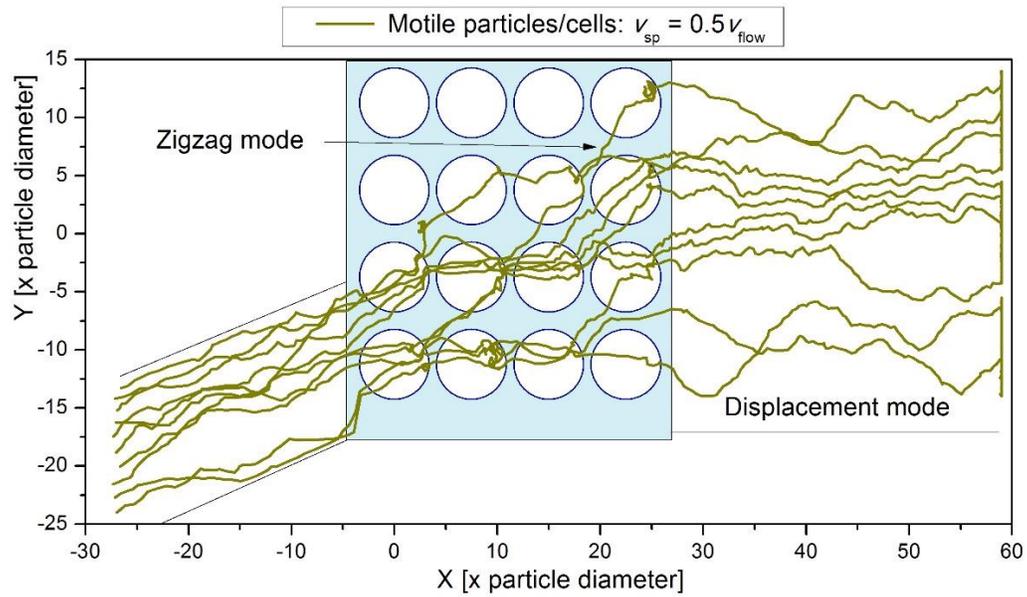

**Fig. 7.** Trajectories of self-propelled particles driven by an external flow with velocity $v_{flow} = 1$ through a 4 x 4 array of attractive parabolic potential energy traps, for an angle, $\gamma = 21.8°$, between the direction of the flow and the array row, and for these values of the self-velocity: $v_0 = 0.25 v_{flow}$ (a) and $v_0 = 0.5 v_{flow}$ (b).

After further parameter optimization and series of simulations (not shown, for brevity), we found a parameter window where the proposed DLD device demonstrated a high efficiency in *separating active particles from passive*.

The case of "clean" separation of self-propelled particles driven by an external flow with velocity $v_{flow} = 1$ through a 4 x 4 array of attractive parabolic potential energy traps, is shown in Fig. 8 for an angle of g = 21.8° between the direction of the flow and the array row, and for self-velocity of active particles: $v_0 = 0.8 v_{flow}$. Indeed, the tracked particle trajectories demonstrate that all the passive particles follow the "displacement" mode and arrive at the band limited by the two lower trap rows, where they can be collected separately from active particles. All these active particles follow the "zigzag" mode and arrive at the band within the upper two trap rows. Thus, the proposed DLD-device that employs an array of attractive traps has been demonstrated to be efficient for the continuous separation of particles (or motile cells like sperm cells) based on their motility.

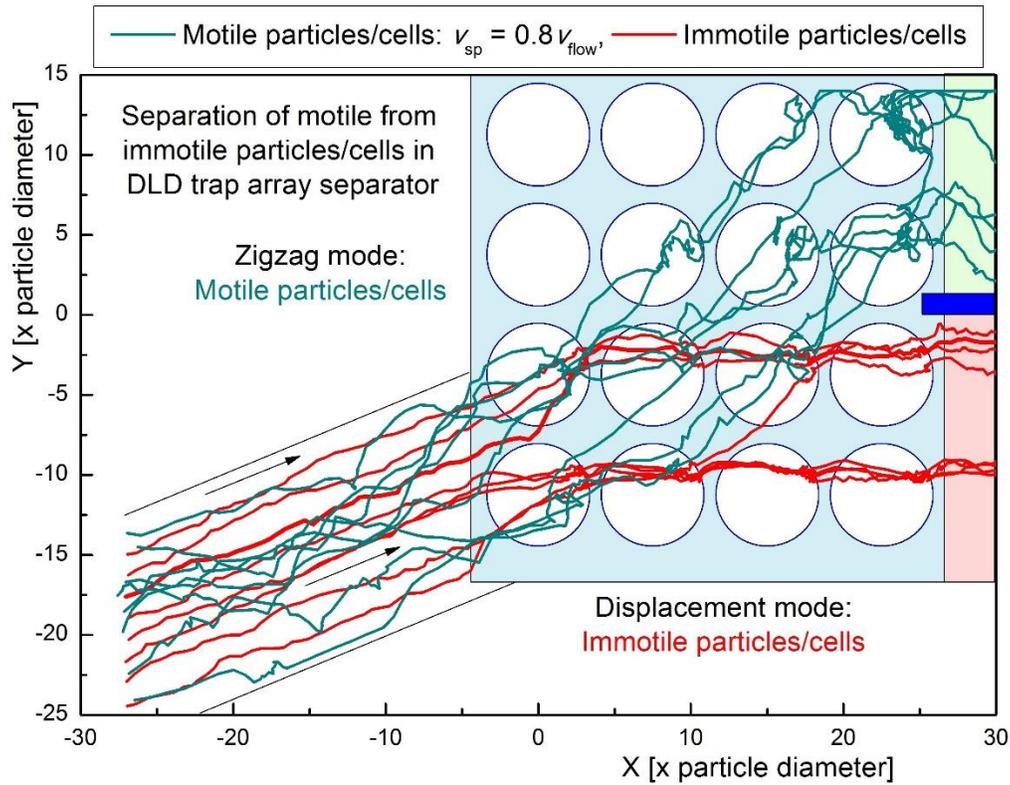

**Figure 8.** Separation of self-propelled particles (motile cells) from passive beads (immotile cells) in a DLD-type separation device formed by a 4 x 4 array of attractive parabolic potential energy traps, for angle of $\gamma = 21.8°$ between the direction of the flow and the array row, and for self-velocity of active particles: $v_0 = 0.8 v_{flow}$.

As compared to the trapping-detrapping separation method, the DLD-type method involves more adjustable parameters and therefore requires a gentle balance between the key parameters of the method, i.e., the angle of the infused flow with respect to the DLD array axis as well as the flow rate and the self-propulsion velocity.

## 5. Discussion and Conclusions

We demonstrated the principles of separation of motile self-propelled micro- and nanoparticles (or motile living cells like bacteria or sperm cells) based on their motility in arrays of attractive potential-energy sites (traps). This has been done in numerical experiments, i.e., in simulations, by solving Langevin-type equations of motion for self-propelled particles and passive beads driven through arrays of attractive sites. The dynamics of driven particles involves self-propulsion (of the active species) and thermal diffusion as well as trapping-detrapping events of the particles in the potential traps.

Two methods of separation were proposed and demonstrated. One method implies selective trapping of passive particles (immotile cells) in an array of traps. The idea of the method is that particles of a mixture of active and passive species being driven via an array of traps become trapped in the attractive sites. Passive particles become located at the bottom of the potential energy traps, which requires higher energy (stronger driving force) to extract them than of active particles that are dynamically located near the edge of the trap, and only a small energy is thus required to extract them from the traps. This "preselection" of active and passive particles already inside the traps determines the selective detrapping of active particles. By applying a weak driving force, that allows to detrap active particles and is insufficient to detrap passive beads, the binary mixture can be effectively separated. As a result, passive particles (or immotile cells) remain trapped in the trap array, while active species (motile cells) undergo the trapping-detrapping dynamics while moving via the array of traps and finally can be collected separately outside the trap.

The second method is somewhat more complex. It implies the selective guiding of passive particles, in a mixture of passive and active particles, via an array of attractive potential sites. The idea of this method could be understood from the energy minimization of the trajectories during the driven motion. This is realized when the trajectories of motion of these passive particles go through the potential-energy minima of the traps and overcome the potential-energy barriers between the traps at the minimal distance between the traps. Clearly, this provides an effective guide or "rail" for the passive particles by the rows of the attractive traps in the array. The condition holds when the driving flow angle is small enough with respect to the direction of the trap rows (for large angles, passive particles can jump to the neighbour trap rows).

On the contrary, the trajectories of the active particles take place at higher potential energies. They go through the edges of the traps, thus near the top of the traps, which is close to the regions between the traps. Therefore, when moving over a trap array, trapping-detrapping events can be considered as small perturbations (provided the self-velocity of the active particles is large enough). As a result, the active particles move on average in the direction of the driving force giving rise to a DLD-type separation: passive particles follow the "displacement" mode, and the active particles execute "zigzag" type motion. This method allows a continuous separation of active and passive particles (or immotile and motile cells) when the different motility particles are collected separately. Note that in this way not only motile particles (cells) can be separated from immotile particles (cells), but also high- from low-motility particles (cells), as demonstrated in our simulations.

The demonstrated methods of separation of active and passive species based on their motility, in arrays of attractive potential energy traps, can be of interest for separating, e.g., high-motility Janus particles needed for various applications like biosensing [41] or mixing in microfluidics or for the elimination of drug-resistant biofilms [13, 14]. On the other hand, selection of high-motility sperm cells is very important for medically assisted methods of artificial fertilization. Our theoretical and numerical predictions open a venue for further research that will be targeted at the experimental verification of the predicted separation methods. As suitable active-passive system, we consider, e.g., motile and immotile sperm cells and synthetic Janus micro- and nanoparticles. Both systems have been previously extensively studied (including related works [7, 8, 11, 12, 33, 35] co-authored by the authors) and these can be employed in the future experiments. In addition, further optimization of the device performance will be of importance for specific systems and conditions. Therefore, further research steps will also include additional simulations for those specific conditions. In terms of the experimental realization of the trap arrays, potential candidates include, e.g., microwells employed for single cell manipulation and isolation (see, e.g., Refs. [42-44]). More advanced and allowing broad tuneability, but at the same time more difficult in realization, can be arrays of rotating flows (vortices), e.g., generated by means of electro-osmotic flows (EOF) [45] or by acoustic excitation in a microfluidic setup [46, 47, 35]. These methods allow to generate well-localized vortices with rather isotropic and fast vortical fluid flow, with the vortex cores directed along the microfluidic channels. In this way, their lateral dimensions are limited by the channel cross-section (hundreds mm) that can accommodate, e.g., two or four vortices. Therefore, the diameters of the vortices can be of tens to hundreds mm. Generation of larger regular arrays of identical vortices (i.e., 4 x 4) is still a challenging task. Vortices, however, possess natural chirality, that results not only in trapping (e.g., micro-magnet generated vortices showing striking capability of trapping particles, like "vortex tweezers", as demonstrated in experiments [48]) but also redirecting the driven particles during the trapping-detrapping events. This in general makes the dynamics of driven particles somewhat more complex and requires further research efforts, both theoretical and experimental.

## 6. Methods

### 6.1. Numerical methods: Molecular-dynamics simulations

The behaviour of the system consisting of motile microswimmers (artificial self-propelled Janus particles or sperm cells), and passive species (synthetic beads or immotile sperm cells and debris) is simulated by numerically integrating the overdamped Langevin equations [6-8, 11, 12, 31, 33-35, 49, 50]:

$$\begin{aligned} \dot{x}_i &= v_0 \cos\theta_i + \xi_{i0,x}(t) + \sum_{ij}^{N} f_{ij,x} + v_{\text{flow},x}, \\ \dot{y}_i &= v_0 \sin\theta_i + \xi_{i0,y}(t) + \sum_{ij}^{N} f_{ij,y} + v_{\text{flow},y}, \\ \dot{\theta}_i &= \xi_{i\theta}(t), \end{aligned} \quad (7)$$

for $i, j$ running from 1 to the total number $N$ of particles, active and passive; $v_0$ is self-velocity of active particles. Here, $x_{i0}(t) = (x_{i0,x}(t), x_{i0,y}(t))$ is a 2D thermal Gaussian noise with correlation functions $\langle x_{0,a}(t)\rangle = 0$, $\langle x_{0,a}(t)x_{0,b}(t)\rangle = 2D_T d_{ab}d(t)$, where $a, b = x, y$ and $D_T$ is the translational diffusion constant of a passive particle at fixed temperature; $x_q(t)$ is an independent 1D Gaussian noise with correlation functions $\langle x_q(t)\rangle = 0$ and $\langle x_q(t)x_q(0)\rangle = 2D_R d(t)$ that models the fluctuations of the propulsion angle $q$.

In simulations, we employ dimensionless units: the unit length is the diameter of the particle, $2r_0$; the damping coefficient, g, is set to unity; the thermal diffusion parameters: temperature, $T = 2.5 \cdot 10^{-3}$, the rotational diffusion coefficient, $D_{rot} = 2.5 \cdot 10^{-2}$. This makes the found relations general and applicable for various motile particles or cells. If needed, these can be easily converted in specific (dimensional) parameters for a particular system. For example, for Janus particles, the diffusion coefficients, $D_T$ and $D_R$, can be directly calculated or extracted from experimentally measured trajectories and mean squared displacement curves (MSD), by fitting to theoretical MSD [50]. Thus, for typical experimental conditions, e.g., a particle with diameter of 2 µm diffusing in water at room temperature, $D_T \approx 0.22$ mm$^2$ s$^{-1}$ and $D_R \approx 0.16$ rad$^2$ s$^{-1}$ [50].

The term, $\sum_{ij}^{N} f_{ij}$, represents, in a compact form, the inter-particle interaction forces in the system, i.e., elastic soft-core repulsive interactions between active particles, between passive beads, and between active and passive particles (for simplicity, it is common (see, e.g., [6, 49]) to present both species by soft interacting disks of radius $r_0$ and repulsive force of modulus $F_{i,j} = k(2r_0 - r_{ij})$ if $r_{ij} < 2r_0$ and $F_{i,j} = 0$ otherwise, where $k$ is the interaction constant); $v_{flow}$ is the flow velocity, and $f_r$ is the acoustic radiation force exerted on the particle. Note that the proportionality coefficient between velocity and force in the overdamped equations (7), i.e., the cumulative damping constant g, is set to unity. It is also assumed that in the overdamped regime the driving force due to the flow is balanced by the Stokes drag, and the net component of velocity of a particle is equal to the fluid flow.

## Acknowledgements

V.R.M. and W.D.M. acknowledge the support of Research Foundation-Flanders (FWO-Vl), Grants No. G029322N, S008423N. V.R.M. acknowledges the support of Research Foundation-Flanders (FWO-Vl), Grant No. K208824N ("Short stay" at RIKEN, Japan). F.N. is supported in part by the Japan Science and Technology Agency (JST) [via the CREST Quantum Frontiers program Grant No. JPMJCR24I2, the Quantum Leap Flagship Program (Q-LEAP), and the Moonshot R&D Grant Number JPMJMS2061], and the Office of Naval Research (ONR) Global (via Grant No. N62909-23-1-2074).